\documentclass[conference]{IEEEtran}
\IEEEoverridecommandlockouts
\usepackage{cite}
\usepackage{amsmath,amssymb,amsfonts}
\usepackage{algorithmic}
\usepackage{graphicx}
\usepackage{textcomp}
\usepackage{xcolor}
\def\BibTeX{{\rm B\kern-.05em{\sc i\kern-.025em b}\kern-.08em
    T\kern-.1667em\lower.7ex\hbox{E}\kern-.125emX}}

\usepackage{caption}
\usepackage{graphicx}
\usepackage{float} 
\usepackage{subcaption}

\begin{document}

\title{Ambiguity Function Analysis of Pilot-Embedded Random OFDM Signals\\

\thanks{This work was supported in part by the Mobile Information Networks-National Science and Technology Major Project under Grant 2025ZD1302000, and in part by the National Natural Science Foundation of China (NSFC) under Grant 62522107 and Grant 62331023. \it(Corresponding author: Fan Liu.)\rm}
}
\iffalse
\author{\IEEEauthorblockN{1\textsuperscript{st} Given Name Surname}
\IEEEauthorblockA{\textit{dept. name of organization (of Aff.)} \\
\textit{name of organization (of Aff.)}\\
City, Country \\
email address or ORCID}
\and
\IEEEauthorblockN{2\textsuperscript{nd} Given Name Surname}
\IEEEauthorblockA{\textit{dept. name of organization (of Aff.)} \\
\textit{name of organization (of Aff.)}\\
City, Country \\
email address or ORCID}
\and
\IEEEauthorblockN{3\textsuperscript{rd} Given Name Surname}
\IEEEauthorblockA{\textit{dept. name of organization (of Aff.)} \\
\textit{name of organization (of Aff.)}\\
City, Country \\
email address or ORCID}
}
\fi

\author{
\IEEEauthorblockN{
Jialin Wu\IEEEauthorrefmark{1}, Fan Liu\IEEEauthorrefmark{1}, Ying Zhang\IEEEauthorrefmark{2}, Yifeng Xiong\IEEEauthorrefmark{3}, Jie Yang\IEEEauthorrefmark{1}, and Shi Jin\IEEEauthorrefmark{1}}
\IEEEauthorblockA{\IEEEauthorrefmark{1}Southeast University, Nanjing, China}
\IEEEauthorblockA{\IEEEauthorrefmark{2}Southern University of Science and Technology, Shenzhen, China}
\IEEEauthorblockA{\IEEEauthorrefmark{3}Beijing University of Posts and Telecommunications, Beijing, China}
\IEEEauthorblockA{Emails: jialinw@seu.edu.cn, fan.liu@seu.edu.cn, zhangying2024@mail.sustech.edu.cn, \\ yifengxiong@bupt.edu.cn, yangjie@seu.edu.cn, jinshi@seu.edu.cn}}

\maketitle


%\maketitle

\begin{abstract}
This paper investigates the statistical ambiguity functions (AFs) of orthogonal frequency division multiplexing (OFDM) waveforms that incorporate deterministic unit-modulus pilot symbols and random data payloads for integrated sensing and communication (ISAC). We derive analytical expressions for the mean squared discrete periodic ambiguity function (DP-AF) and fast-slow-time ambiguity function (FST-AF) of such pilot-embedded OFDM signals. Our analysis demonstrates that, under a fixed signal length and constellation scheme, the mean squared DP-AF depends jointly on the pilot patterns, pilot symbols and number of pilots, while the mean squared FST-AF relies only on the number of pilots. Numerical simulations closely match the theoretical expressions. Furthermore, in numerical results, we show that different pilot patterns correspond to DP-AF with distinct characteristics, offering relevant considerations for pilot design in communication-centric ISAC systems.
\end{abstract}


\section{Introduction}
In the imminent deployment of sixth generation (6G) technology, the dual need for robust connectivity and precise environmental awareness is driving the evolution of integrated sensing and communication (ISAC), a key 6G paradigm that merges the functions of communication and sensing on a shared platform, gaining strong interest for its advantages in improved resource utilization and mutual performance enhancement \cite{6GBeyond}. %\cite{Seventy_Years}

To fully leverage these benefits, a key task in ISAC is to design dual-functional waveforms that jointly serve communication and sensing purposes. Current efforts primarily follow three distinct directions: communication-centric, sensing-centric and joint design schemes \cite{Over_the_Years}. Among these, communication-centric schemes adapt conventional communication waveforms for sensing functions, which offer notable practical benefits for implementation owing to their alignment with existing communication network standards.

Current communication-centric methods support sensing primarily through structured reference signals \cite{RSWCNC}. While these deterministic sequences offer favorable correlation properties for basic parameter estimation, their confined time-frequency allocation restricts the achievable sensing performance \cite{Liu2026JSAC}. This constraint underscores the necessity of additionally harnessing the remaining resources, namely random data payloads, to improve sensing accuracy and spectral efficiency without degrading communication performance. % raderconfprs, RSVTC

% 此处添加DRT相关内容：有必要研究用随机通信数据进行感知的感知性能分析

Towards that end, recent studies have begun to examine how to design the transmission of random communication signals to improve sensing performance \cite{CP-OFDM}, \cite{iceberg}. Orthogonal frequency division multiplexing (OFDM), as the commonly used waveform standard in contemporary systems, has been a primary focus of such investigations. The authors of \cite{CP-OFDM} derived closed-form expressions of the mean squared auto-correlation function (ACF) to evaluate the ranging performance of different orthogonal waveforms with random data symbols. It is demonstrated that among all orthogonal waveforms with cyclic prefix (CP), OFDM achieves minimal sidelobes in constellation schemes such as quadrature amplitude modulation (QAM) or phase shift keying (PSK). Expanding on this, \cite{iceberg} incorporated pulse shaping into the analysis of ACF. For QAM or PSK constellation schemes, it is found that OFDM again attains minimal sidelobes under Nyquist pulse shaping. These studies collectively demonstrate the great potential of adopting OFDM as the ranging waveform under random signaling in communication-centric ISAC systems.

While the ACF, corresponding to the zero-Doppler cross-section of the ambiguity functions (AFs), is a fundamental tool for evaluating ranging performance, comprehensive sensing in dynamic environments requires joint analysis in both delay and Doppler dimensions. The AFs serve as the fundamental tool for this purpose \cite{AndrewAF}, characterizing the response of a waveform to matched filtering in the delay-Doppler domain. Extending the evaluation to this domain, \cite{zhang2025discrete} developed a unified characterization of the discrete periodic ambiguity function (DP-AF) and the fast-slow-time ambiguity function (FST-AF) for random waveforms. This analysis reveals that under the DP-AF framework, no waveform can achieve minimum expected sidelobe level (ESL) over a two-dimensional compact region in the delay-Doppler domain. In contrast, under the FST-AF framework, which is essentially an approximation of the DP-AF in the low-Doppler regime, optimality becomes constellation-dependent: OFDM minimizes sidelobes for sub-Gaussian distributions, while the orthogonal time frequency space (OTFS) becomes optimal for super-Gaussian ones. 

While the aforementioned investigations have systematically analyzed the statistical ambiguity properties of OFDM under random ISAC signaling, they have not yet extended their research to practical communication frame structures with hybrid composition, which consist of both random data payloads and deterministic pilot sequences \cite{Both_Pilots_and_Data_Payloads}. The sensing performance of a joint processing scheme that coherently combines both pilots and data remains to be further evaluated. Furthermore, the impact of pilot sequences and their placement patterns on the overall sensing ambiguity properties have not been adequately explored. These limitations motivate us to investigate pilot-embedded random OFDM signals, particularly their statistical ambiguity properties.

To bridge this research gap, this paper investigates the discrete AFs of random OFDM signals with embedded unit-modulus pilots. Analytical formulas for the mean squared DP-AF and FST-AF are derived. Our analysis demonstrates that, given a fixed signal length and constellation scheme, the mean squared DP-AF varies with the pilot patterns, pilot symbols, and the number of pilots, while the mean squared FST-AF depends exclusively on the number of pilots. Numerical results show strong consistency between theoretical and simulated outcomes. Additionally, in numerical results, we show that different pilot patterns correspond to DP-AF with distinct characteristics, offering relevant considerations for pilot design in communication-centric ISAC systems.

\it Notations\rm: We use capital letters denoted in bold (e.g., $\mathbf{X}$) for matrices and lowercase letters denoted in bold (e.g., $\mathbf{s}$) for vectors. Scalars are written in a normal font (e.g., $L$). The $n$-th element of a vector $\mathbf{p}$ is written as $p_n$. The element in the $m$-th row and $n$-th column of a matrix $\mathbf{H}$ is denoted by a $h_{m,n}$. The operator $\operatorname{vec}(\cdot)$ signifies vectorization. $\otimes$ represents the Kronecker product. The symbols $(\cdot)^H$, $(\cdot)^T$ and $(\cdot)^*$ represent the Hermitian transpose, transpose, and complex conjugate, respectively. The expectation operator is represented by $\mathbb{E}(\cdot)$. The Kronecker delta $\delta_{m,n}$ equals $1$ if $m=n$ and $0$ otherwise, $\delta_{l,r,m,n}$ equals $1$ if $l=r=m=n$ and $0$ otherwise. $\mathbf{F}_N$ denotes the $N\times N$ normalized discrete Fourier transform matrix. The modulo $N$ operation is denoted by $\langle\cdot\rangle_N$. For a matrix $\mathbf{X}$, its entry-wise squared magnitude is given by $\left|\mathbf{X}\right|^2$.
\iffalse
\begin{align}\nonumber\delta_{m,n}=
\begin{cases}
0, & m\neq n; \\
1, & m=n.  
\end{cases}\end{align}\fi

\section{Signal Model and Sensing Performance Evaluation}
\subsection{Pilot-embedded OFDM Signal Model}
This section presents two kinds of OFDM signal models with embedded pilots. While the one-dimensional (1D) model characterizes the time-domain samples transmitted in sequence, the two-dimensional (2D) model naturally captures the structure of OFDM signals mapped across both subcarriers and symbols in the time-frequency grid.

\subsubsection{1D Signaling}
We consider the scenario where $N$ symbols are transmitted. The transmitted symbols are denoted as $\mathbf{s}= \begin{bmatrix} s_0,s_1,\ldots,s_{N-1} \end{bmatrix}^T\in\mathbb{C}^N$, which is composed of $L$ pilot symbols and $(N - L)$ i.i.d. communication symbols. The symbols used for pilots are deterministic. The communication symbols are randomly drawn from a constellation, which is considered for being subject to the following constraints:
%(\it Unit Power and Rotational Symmetry\rm)

\textbf{Assumption 1} : \it For a random communication symbol $s_{c}$, we assume that its statistical properties are characterized by unit power, zero mean, and zero pseudo-variance \rm\cite{zhang2025discrete}\it, namely 
\begin{equation}\label{eq:asp1}\mathbb{E}(\left|s_{c}\right|^2)=1,\quad\mathbb{E}(s_{c})=0,\quad\mathbb{E}(s_{c}^2)=0.\end{equation}
And we additionally assume 
\begin{equation}\label{eq:asp2}\mathbb{E}(\left|s_{c}\right|^2s_{c})=\mathbb{E}(\left|s_{c}\right|^2s_{c}^*)=0,
\end{equation}
which holds for most symmetric constellations. \rm

The kurtosis of the constellation, which plays a fundamental role in determining the AFs of random communication signals \cite{zhang2025discrete}, is defined as:
\begin{equation}\kappa:=\frac{\mathbb{E}\left\{\left|s_{c}-\mathbb{E}\left(s_{c}\right)\right|^{4}\right\}}{\mathbb{E}^{2}\left\{\left|s_{c}-\mathbb{E}\left(s_{c}\right)\right|^{2}\right\}}=\mathbb{E}\left\{\left|s_{c}\right|^{4}\right\}.\end{equation}

The subcarrier indices of the $L$ pilots are described by a set $I=\{i_0,i_1,\ldots,i_{L-1}\}$, which constitutes the pilot pattern. The symbol vector $\mathbf{s}$ can then be partitioned into a data component $\mathbf{d}$ and a pilot component $\mathbf{p}$, whose elements satisfy:
\begin{align}\label{eq:defdi}d_i =
\begin{cases} 
0, &\text{if}\ i \in I\\
s_i, &\text{else}
\end{cases},\ \ p_i =
\begin{cases}
s_i, &\text{if}\ i \in I\\
0, &\text{else}
\end{cases}.
\end{align}

To encode the locations of the pilots, a binary indicator vector $\mathbf{w}$ is introduced, where the $i$-th element is defined as $w_i=1$ if $i \in I$, and conversely, $w_i=0$ if $i \notin I$. It follows that $\mathbf{s} = \mathbf{d} + \mathbf{p}$, and
\begin{equation}\label{eq:swpd}
s_i = w_i \cdot p_i+(1-w_i)\cdot d_i,\ \forall \ 0\le i\le N-1.
\end{equation}

Furthermore, we consider the case where the energy of the deterministic pilot symbols equals to that of the communication symbols, i.e., $\left|s_i\right|^{2}=1$, if $\ i \in I$. This equal-power assumption is made to facilitate a tractable analysis and a fair comparison. In this work, we refer to pilots satisfying this condition as unit-modulus pilots. 

Under the above assumptions, the pilot-embedded 1D OFDM signal is modeled as:
\begin{equation}\label{eq:1dsignal}
\mathbf{x}=\mathbf{F}_N^H\cdot\mathbf{s}=\mathbf{F}_N^H\cdot\mathbf{d}+\mathbf{F}_N^H\cdot\mathbf{p}.
\end{equation}

\subsubsection{2D Signaling}
We now discuss a general two-dimensional modulation scheme that can be expressed in the form of a time-domain OFDM signal matrix:
\begin{equation}\label{eq:2dsignal}
\mathbf{X}=\mathbf{F}_N^H\cdot\mathbf{S}\cdot\mathbf{I}_M,
\end{equation}
where $\mathbf{S}\in \mathbb{C}^{N\times M}$ is composed of $L$ unit-modulus pilot symbols and $(MN - L)$ i.i.d. communication symbols. The communication symbols are randomly drawn from a constellation, which is considered adopting the same assumptions as in 1D signaling. 

The OFDM signal $\mathbf{X}$ is represented as an $N\times M$ matrix, where each row corresponds to a fast-time sample and each column to a slow-time sample, thereby occupying $N$ slots in the fast-time dimension and $M$ slots in the slow-time dimension.

\subsection{Sensing Performance Metrics}
In this subsection, we introduce the discrete ambiguity functions as the sensing performance metrics, which are derived in \cite{zhang2025discrete} for 1D and 2D signals.

\subsubsection{DP-AF of the 1D OFDM Signal}
%Shifting the signal in time by $k$ samples and applying a Doppler frequency shift indexed by $q$ can be expressed using matrix multiplication.
Consider a discrete-time signal with length $N$. Shifting the signal in time by $k$ samples can be expressed using matrix multiplication. Specifically, the insertion and removal of CP give rise to the periodic time-shift matrix, which is denoted as 
\begin{equation}\mathbf{J}_{N,k}:=\begin{bmatrix}
\mathbf{0} & \mathbf{I}_k \\
\mathbf{I}_{N-k} & \mathbf{0}
\end{bmatrix}.\end{equation}
Additionally, a Doppler frequency shift corresponding to index $q$ can be expressed as a frequency-domain shifting matrix, which is given by
\begin{equation}\mathbf{D}_{N,q}:=\mathrm{Diag}\left(1,e^{j\frac{2\pi q}{N}},\ldots,e^{j\frac{2\pi q(N-1)}{N}}\right).\end{equation}

Assume that the time-domain signal $\mathbf{x}=\mathbf{F}_N^H\cdot\mathbf{s}$ undergoes CP insertion followed by propagation through a multi-target channel. Each target is associated with a delay index $k_u$, a Doppler index $q_u$ and a complex amplitude $\beta_u$, $u=1,2,\ldots,U$. Once the receiver strips the CP, the received echo can be modeled as \cite{zhang2025discrete}
\begin{equation}\mathbf{y}=\sum_{u=1}^U\beta_u\mathbf{D}_{N,q_u}\mathbf{J}_{N,k_u}\mathbf{x}+\mathbf{z},\end{equation}
where $\mathbf{z}$ represents the AWGN. The matched-filter output, obtained by correlating $\mathbf{y}$ with a time-frequency shifted version of $\mathbf{x}$, is given by:
\begin{align}\nonumber&\ \tilde{y}_{k,q}^{\mathrm{MF}}=\mathbf{x}^H\mathbf{J}_{N,k}^T\mathbf{D}_{N,q}^*\mathbf{y}\\&=\sum_{u=1}^U\beta_u e^{\frac{j2\pi k_u(q-q_u)}{N}}\mathbf{x}^H\mathbf{J}_{N,k-k_u}^T\mathbf{D}_{N,q-q_u}^*\mathbf{x}+\tilde{z}_{k,q},
\end{align}
where $\tilde{z}_{k,q}=\mathbf{x}^H\mathbf{J}_{N,k}^T\mathbf{D}_{N,q}^*\mathbf{z}$, and we accordingly define the DP-AF as
\begin{align}\mathcal{A}_{\mathrm{DP}}(k,q)=\mathbf{x}^H\mathbf{J}_{N,k}^T\mathbf{D}_{N,q}^*\mathbf{x},\quad(k,q)\in\mathbb{Z}_N^2,\end{align}
where $k$ is the delay index, and $q$ is the Doppler index. To enable effective multi-target detection, the squared matched-filter output $|\tilde{y}_{k,q}^{\mathrm{MF}}|^2$ should display sharp and distinct peaks at the coordinates $(k,q)=(k_u,q_u)$, while maintaining low sidelobe values elsewhere. This behavior is largely determined by the overall form of $\mathcal{A}_{\mathrm{DP}}(k,q)$, which leads to the definition of the sidelobe level \cite{zhang2025discrete}:
\begin{align}\label{eq:def_dp}
|\mathcal{A}_{\mathrm{DP}}(k,q)|^2 
 & =|\mathbf{x}^H\mathbf{D}_{N,q}\mathbf{J}_{N,k}\mathbf{x}|^2,\quad(k,q)\neq(0,0),
\end{align}
and the level of mainlobe is given by $|\mathcal{A}_{\mathrm{DP}}(0,0)|^2=|\mathbf{x}^H\mathbf{x}|^2$.

\subsubsection{FST-AF of the 2D OFDM Signal}
We consider the transmission of the ISAC signal $\operatorname{vec}\left(\mathbf{X}\right)$ with size $MN$. Under a sufficiently small Doppler frequency, the phase shift is assumed invariant across blocks of $N$ samples. This facilitates the segmentation of the signal $\operatorname{vec}\left(\mathbf{X}\right)$ into $M$ column vectors, thereby establishing a two-dimensional fast-slow-time (FST) representation $\mathbf{X}\in\mathbb{C}^{N\times M}$. Additionally, to accommodate the maximum target delay, a CP is incorporated at the transmitter for each slow-time block.
In the multi-target scenario, the received signal following CP removing can be approximated as \cite{zhang2025discrete}
\begin{equation}\mathbf{y}\approx\sum_{u=1}^U\beta_u\left(\mathbf{D}_{N,q_u}\otimes\mathbf{J}_{N,k_u}\right)\mathbf{x}+\mathbf{z}.\end{equation}
The matched-filter output is:
\begin{align}
 & \nonumber\ \tilde{y}_{k,q}^{\mathrm{MF}}=\mathbf{x}^{H}\left(\mathbf{D}_{M,q}^{*}\otimes\mathbf{J}_{N,k}^{T}\right)\mathbf{y} \\
 & \approx\sum_{u=1}^{U}\beta_{u}\mathbf{x}^H\left(\mathbf{D}_{M,q-q_u}^*\otimes\mathbf{J}_{N,k-k_u}^T\right)\mathbf{x}+\tilde{z}_{k,q},
\end{align}
where $\tilde{z}_{k,q}=\mathbf{x}^H\left(\mathbf{D}_{M,q}^*\otimes\mathbf{J}_{N,k}^T\right)\mathbf{z}$, and the FST-AF is accordingly defined as
\begin{equation}\mathcal{A}_{\mathrm{FST}}\left(k,q\right):=\mathbf{x}^H\left(\mathbf{D}_{M,q}^*\otimes\mathbf{J}_{N,k}^T\right)\mathbf{x},\quad k\in\mathbb{Z}_N,q\in\mathbb{Z}_M,\end{equation}
which can be represented in a matrix form \cite{zhang2025discrete}:
\begin{align}\mathcal{A}_\mathrm{FST}&=\sqrt{MN}\mathbf{F}_N^H\left|\mathbf{F}_N\mathbf{X}\right|^2\mathbf{F}_M.
\end{align}

The FST-AF of the 2D OFDM signal $\mathbf{X}=\mathbf{F}_N^H\cdot\mathbf{S}\cdot\mathbf{I}_M$ can be further expressed as
\begin{align}\label{eq:fst}\mathcal{A}_\mathrm{FST}&=\sqrt{MN}\mathbf{F}_N^H\left|\mathbf{S}\right|^2\mathbf{F}_M.
\end{align}

\section{Statistical Characterization of the AFs for Pilot-Embedded Random OFDM Signals}
Since the pilot-embedded OFDM signals in \eqref{eq:1dsignal} and \eqref{eq:2dsignal} contain random communication symbols, the DP-AF and FST-AF of these signals are random functions. Therefore, it is necessary to study the statistical characteristics of DP-AF and FST-AF. Specifically, the expected sidelobe level (ESL) and the expected integrated sidelobe level (EISL) \cite{zhang2025discrete} are important metrics for evaluating the capability of multi-target resolution reflected in a discrete ambiguity function $\mathcal{A}(k,q)$, which are defined as:
\begin{align}
\mathrm{ESL}=\mathbb{E}(|\mathcal{A}(k,q)|^2),\quad\forall(k,q)\neq(0,0),
\end{align}
\begin{align}
\mathrm{EISL}=\sum_{k=0}^{N-1}\sum_{q=0}^{N-1}\mathbb{E}(|\mathcal{A}(k,q)|^2)-\mathbb{E}(|\mathcal{A}(0,0)|^2).
\end{align}

In this section, we presents the derivation of the analytical expressions of ESL and EISL for pilot-embedded random OFDM signals.

\subsection{Statistical Characterization of the DP-AF}
\textbf{Proposition 1}. \it The mean squared DP-AF for random OFDM signals with embedded unit-modulus pilots is \rm
\begin{align}\ &\ \nonumber\mathbb{E}\left(|\mathcal{A}_{\mathrm{DP}}(k,q)|^2\right)\\&=\begin{cases} N^2+(\kappa-1)(N-L), \quad & \text{if}\ k=0,\ q=0,\\
(\kappa-1)(N-L), \quad &  \text{if}\ k\neq0,\ q=0,\\
f_{I}(k,q), \quad &\text{else},\end{cases}\end{align}
\iffalse
\begin{equation}\begin{aligned}&\ \mathbb{E}\left(|\mathcal{A}_{\mathrm{DP}}(k,q)|^2\right)\\&=N^2\delta_{k,0}\delta_{q,0}+\left[(\kappa-2)(N-L)-L^2\right]\delta_{q,0} + f_{I}(k,q).\end{aligned}\end{equation}
\fi
\it where $(k,q)\in\mathbb{Z}_N^2$, $f_{I}(k,q)$ is a function changing with pilot patterns and pilot symbols: 
\begin{equation}\begin{aligned}
&\ f_{I}(k,q) %= &\left|\mathbf{p}^\top\mathbf{F}_N^H\mathbf{D}_{N,q}^*\mathbf{J}_{N,k}^*\mathbf{F}_N\mathbf{p}^*\right|^2+N-\mathbf{w}^\top\mathbf{F}_N^H\mathbf{D}_{N,q}^*\mathbf{F}_N\mathbf{w}\\
%\\&=\left|\sum_{n=0}^{N-1}p_np_{\langle n-q\rangle_N}^*\cdot e^{j2\pi nk/N}\right|^2+N-\sum_{n=0}^{N-1}w_nw_{\langle n-q\rangle_N}
\\&=\left|\mathbf{p}^H\mathbf{F}_N\mathbf{D}_{N,q}\mathbf{J}_{N,k}\mathbf{F}_N^H\mathbf{p}\right|^2+N-\mathbf{w}^H\mathbf{F}_N\mathbf{D}_{N,q}\mathbf{F}_N^H\mathbf{w}.
%=&\left|\mathbf{p}^\top\mathbf{D}_{N,k}\mathbf{J}_{N,q}\mathbf{p}^*\right|^2+N-\mathbf{w}^\top\mathbf{J}_{N,q}\mathbf{w}
\end{aligned}\end{equation}
Proof. \rm Please refer to Appendix A. \hfill$\blacksquare$
\iffalse
\textbf{Proposition 2} (Invariance of EISL). \it For all pilot patterns, the normalized $\mathrm{EISL_{DP}}$ is a constant value $N-1$. 
\begin{align}
\frac{\sum_{k=0}^{N-1}\sum_{q=0}^{N-1}\mathbb{E}(\left|\mathcal{A}_{\mathrm{DP}}(k,q)\right|^2)-\mathbb{E}(\left|\mathcal{A}_{\mathrm{DP}}(0,0)\right|^2)}{\mathbb{E}(\left|\mathcal{A}_{\mathrm{DP}}(0,0)\right|^2)} & =N-1
\end{align}
Proof. \rm 
\fi

The mean squared zero-Doppler cross-section is
\begin{equation}\ \mathbb{E}\left(|\mathcal{A}_{\mathrm{DP}}(k,0)|^2\right)=N^2\delta_{k,0}+(\kappa-1)(N-L).\end{equation}
\iffalse
The mean squared zero-delay slice is
\begin{equation}\ \mathbb{E}\left(|\mathcal{A}_{\mathrm{DP}}(0,q)|^2\right)=\begin{cases} N^2+(\kappa-1)(N-L), \ \text{if}\ q=0\\
f_{I}(0,q),\ \text{else}\end{cases}\end{equation}
\fi

When $q=0$, the ESL of DP-AF is $(\kappa-1)(N-L)$,  and when $q\neq0$, the ESL of DP-AF is $f_{I}(k,q)$. It is noteworthy that, under the DP-AF formulations, sidelobes in regions where the Doppler index $q\neq0$ vary with the pilot symbols and their patterns. From the results in \cite{zhang2025discrete}, the volume of the DP-AF is
\begin{align}\sum_{k=0}^{N-1}\sum_{q=0}^{N-1}\left|\mathcal{A}_{\mathrm{DP}}(k,q)\right|^2=N\|\mathbf{x}\|_2^4=N\left|\mathcal{A}_{\mathrm{DP}}(0,0)\right|^2,
\end{align}
Then the EISL under the DP-AF formulations is given by:
\begin{align}
\mathrm{EISL_{DP}}& \nonumber=(N-1)\mathbb{E}(|\mathcal{A}_{\mathrm{DP}}(0,0)|^2)\\& =(N-1)\left[N^2+(\kappa-1)(N-L)\right].
\end{align}

Therefore, under a fixed signal length and constellation scheme, $\mathrm{EISL_{DP}}$ remains invariant when the number of pilots $L$ remains constant.

\subsection{Statistical Characterization of the FST-AF}
\textbf{Proposition 2}. \it The mean squared FST-AF for random OFDM signals with embedded unit-modulus pilots is
\begin{align}\label{eq:expect_fst}&\ \nonumber\mathbb{E}\left(|\mathcal{A}_{\mathrm{FST}}(k,q)|^2\right)\\&=L+(MN-L)\cdot \kappa+M^2N^2\delta_{k,0}\delta_{q,0}-MN,\end{align}
\ \ \ \ \ \ where $k\in\mathbb{Z}_N$ and $q\in\mathbb{Z}_M$.

Proof. \rm Please refer to Appendix B.\hfill$\blacksquare$

The ESL of FST-AF can be expressed as
\begin{equation}\ \mathrm{ESL_{FST}}=L+(MN-L)\cdot \kappa-MN.\end{equation}

It is noteworthy that, under the FST-AF formulations, the levels of sidelobe all equal to the same value and do not vary with the pilot symbols and their patterns. The EISL under the FST-AF formulations is given by:
\begin{align}
\mathrm{EISL_{FST}} & =(MN-1)\left[L+(MN-L)\cdot \kappa-MN\right],
\end{align}
which depends solely on the number of pilots, the constellation schemes, and the signal length.

\section{Numerical Results}
This section presents three-dimensional plots of the theoretical and simulated mean squared AFs, comparing signals with and without pilots to examine the impact of pilots. For 1D OFDM signals, we additionally demonstrate the impact of pilot patterns on their DP-AF. Moreover, 16-QAM is selected as the constellation, and the corresponding $\kappa=1.32$. Each simulated result is obtained by averaging over $1000$ independent Monte‑Carlo trials. The close alignment between simulated and theoretical curve validates the theoretical derivations.

\begin{figure}[htbp]
    \raggedright
    \ \ \begin{subfigure}{.47\linewidth}
		\includegraphics[width=.3\linewidth]{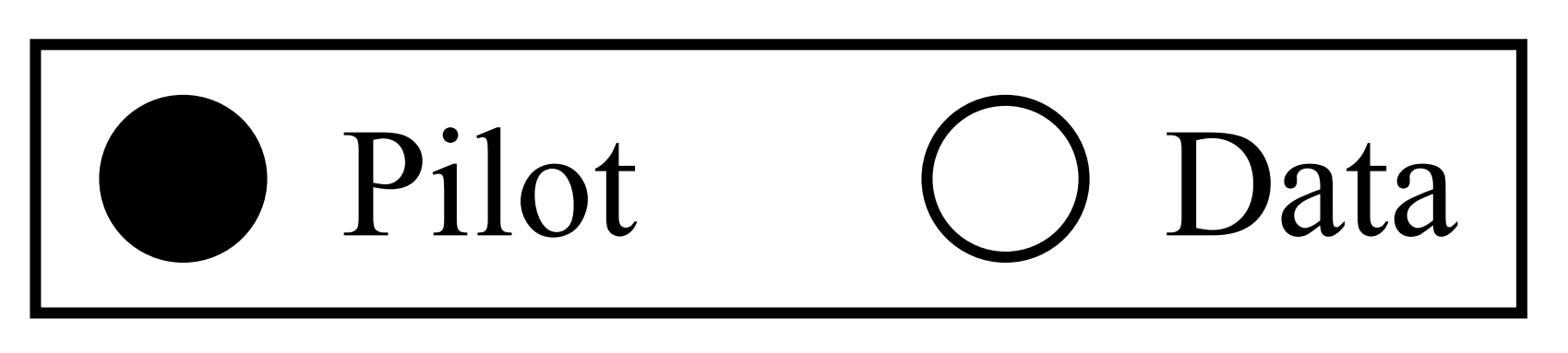}
		\label{chutian3}%文中引用该图片代号
	\end{subfigure}\\
	\centering
	\begin{subfigure}{.47\linewidth}
		\centering
		\includegraphics[width=1\linewidth]{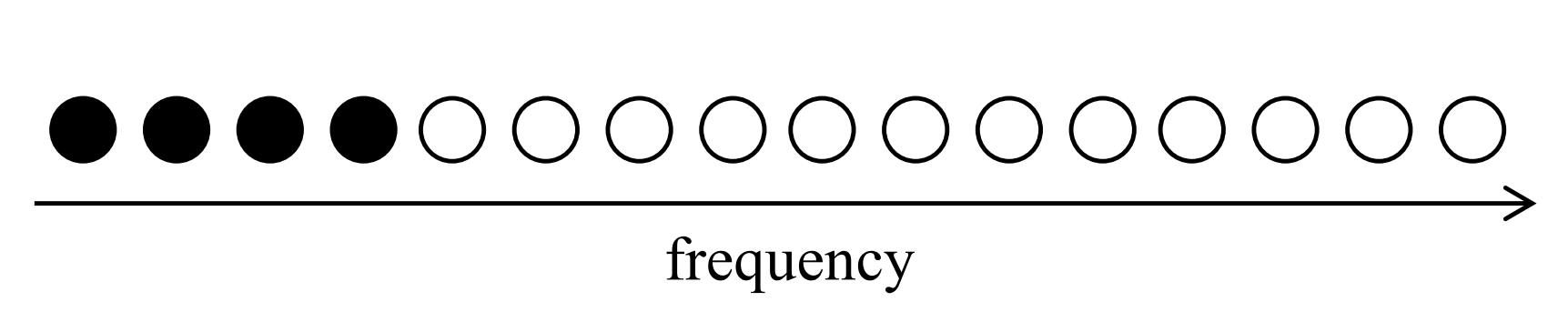}
		\caption{Block-type}
		\label{chutian3}%文中引用该图片代号
	\end{subfigure}
    \centering
	\begin{subfigure}{.47\linewidth}
		\centering
		\includegraphics[width=1\linewidth]{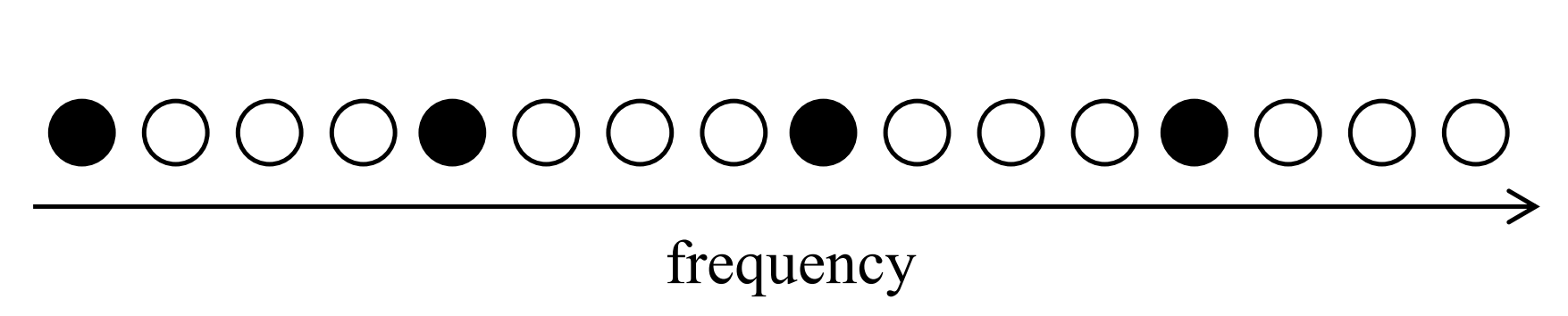}
		\caption{Comb-type}
		\label{chutian3}%文中引用该图片代号
	\end{subfigure}
	\caption{Two different types of pilot patterns.} 
	\label{fig3}
\end{figure}

\begin{figure}[htbp]
	\centering
	\begin{subfigure}{0.48\linewidth}
		\centering
		\includegraphics[width=1\linewidth]{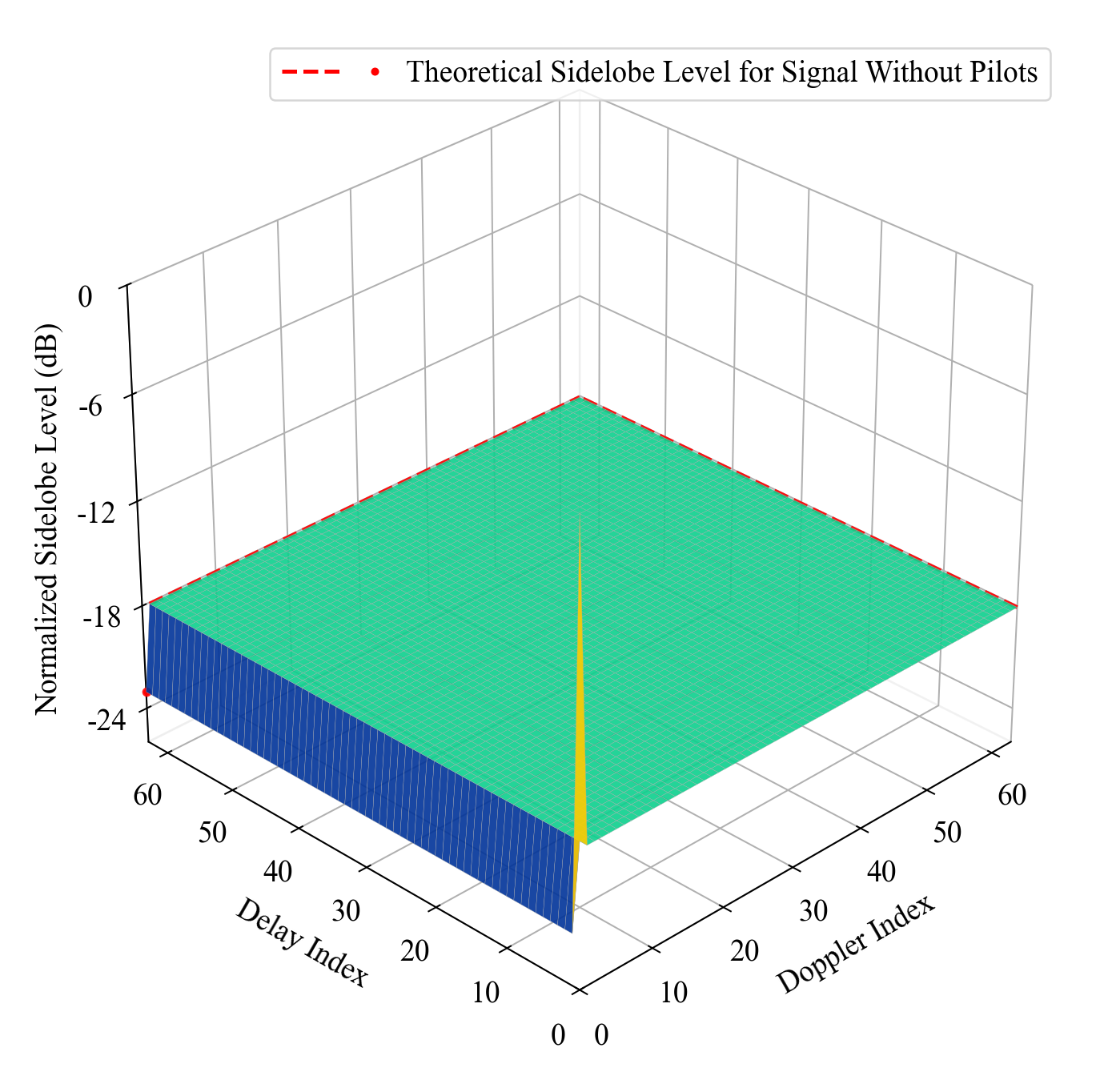}
		\caption{Without Pilot, Theoretical}
		\label{chutian3}%文中引用该图片代号
	\end{subfigure}
	\centering
	\begin{subfigure}{0.48\linewidth}
		\centering
		\includegraphics[width=1\linewidth]{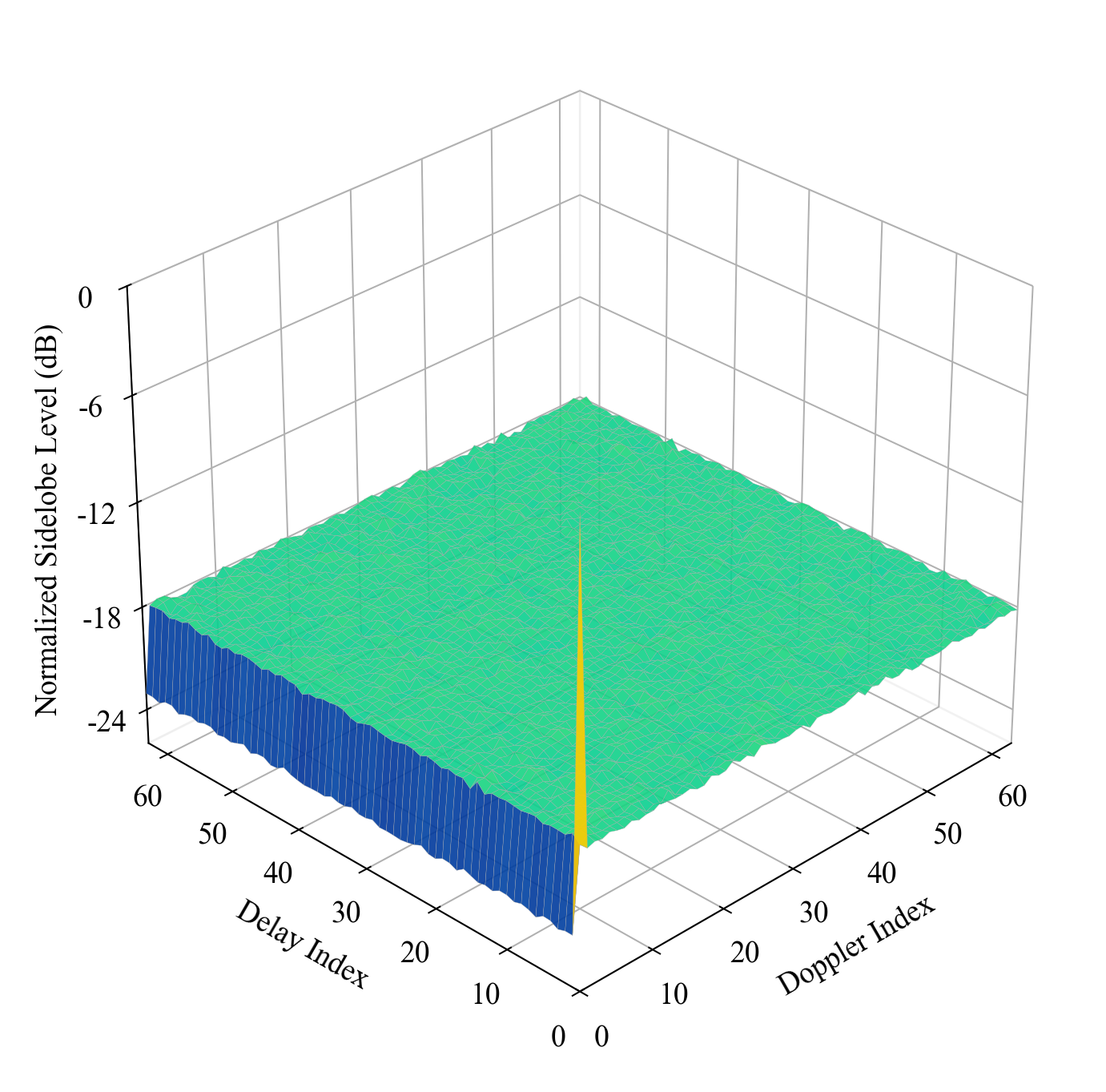}
		\caption{Without Pilot, Simulated}
		\label{chutian3}%文中引用该图片代号
	\end{subfigure}
	\centering
	\begin{subfigure}{0.48\linewidth}
		\centering
		\includegraphics[width=1\linewidth]{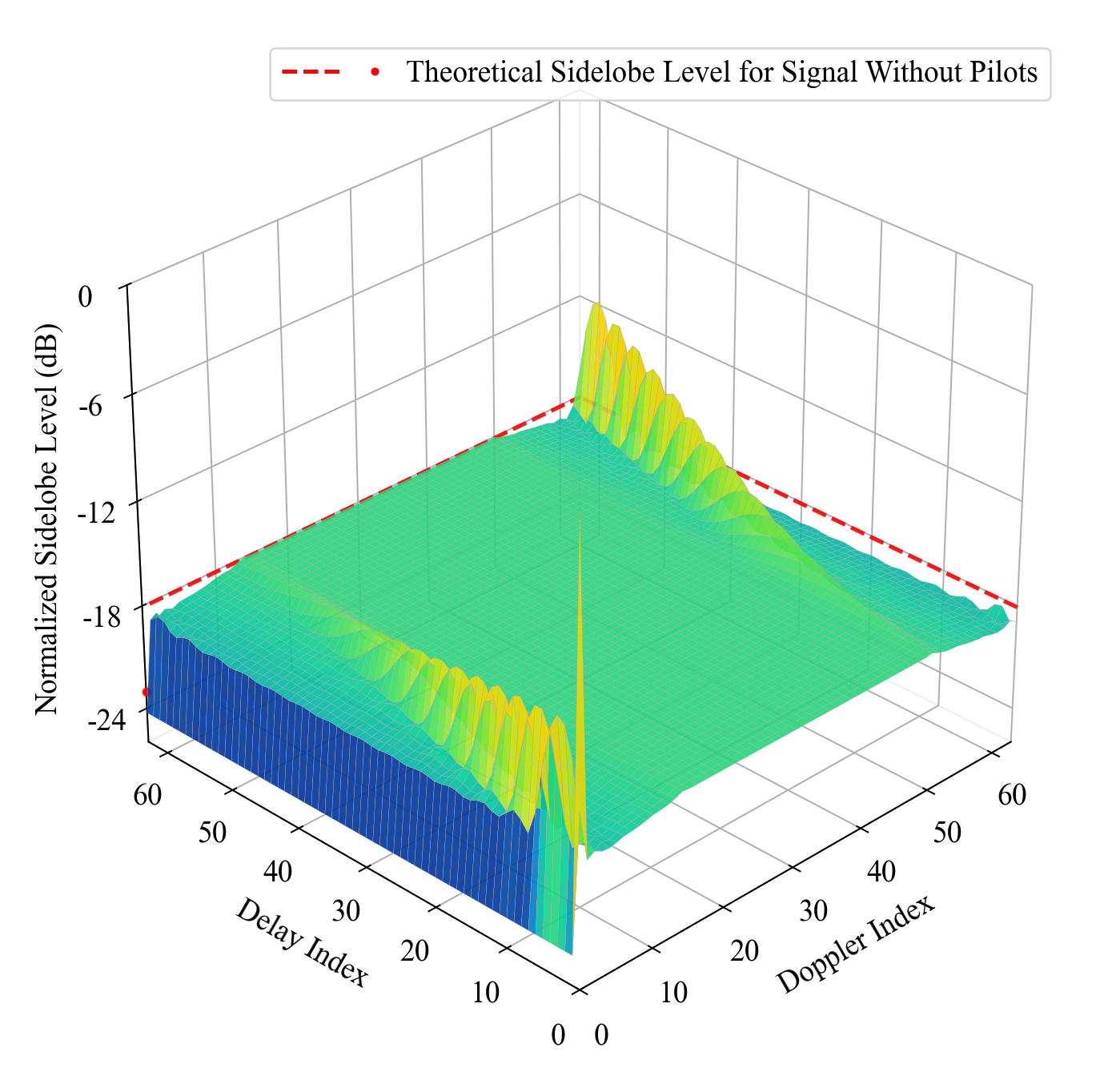}
		\caption{Block-type Pilot, Theoretical}
		\label{chutian3}%文中引用该图片代号
	\end{subfigure}
	\centering
	\begin{subfigure}{0.48\linewidth}
		\centering
		\includegraphics[width=1\linewidth]{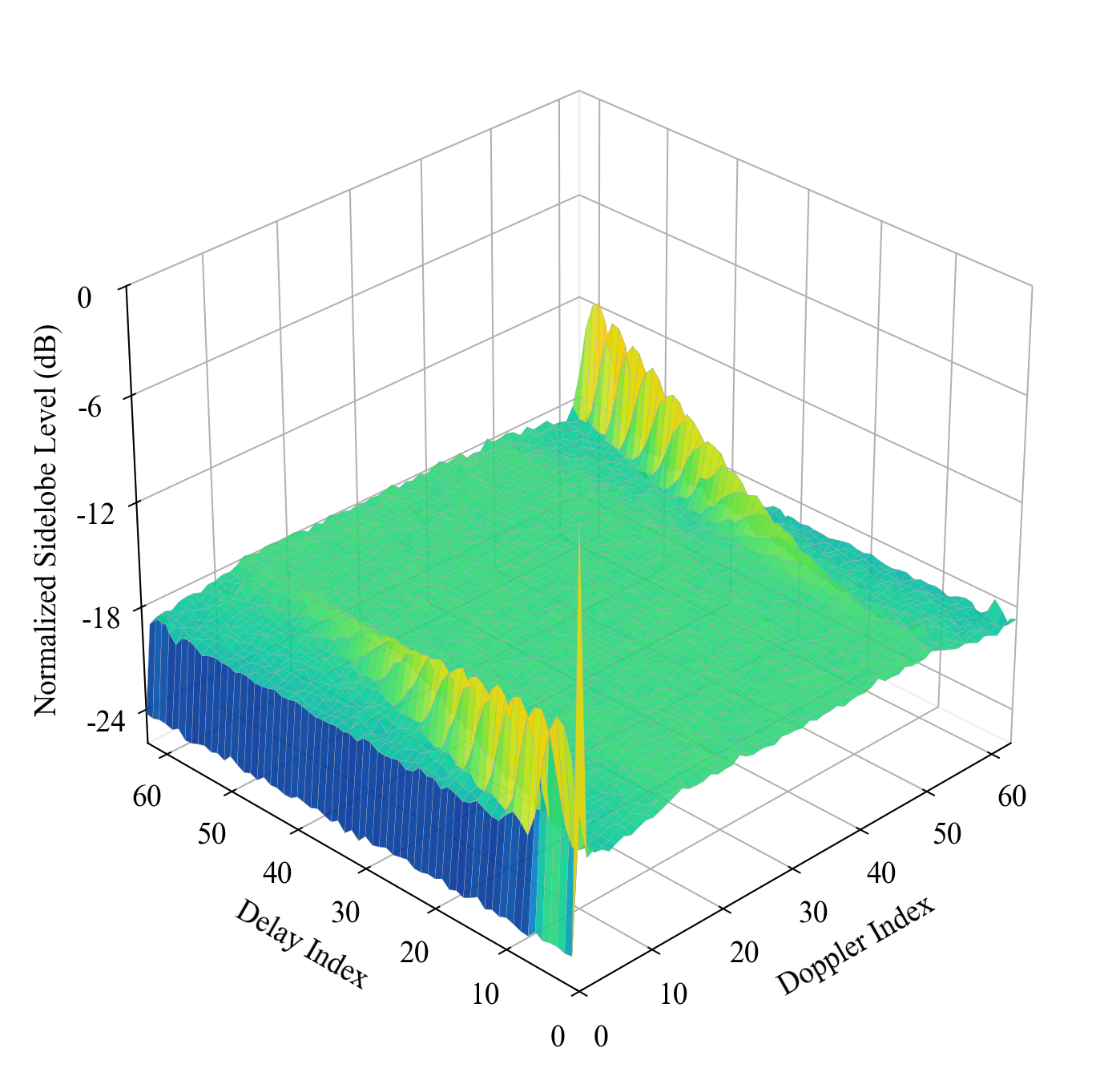}
		\caption{Block-type Pilot, Simulated}
		\label{chutian3}%文中引用该图片代号
	\end{subfigure}
    \centering
	\begin{subfigure}{0.48\linewidth}
		\centering
		\includegraphics[width=1\linewidth]{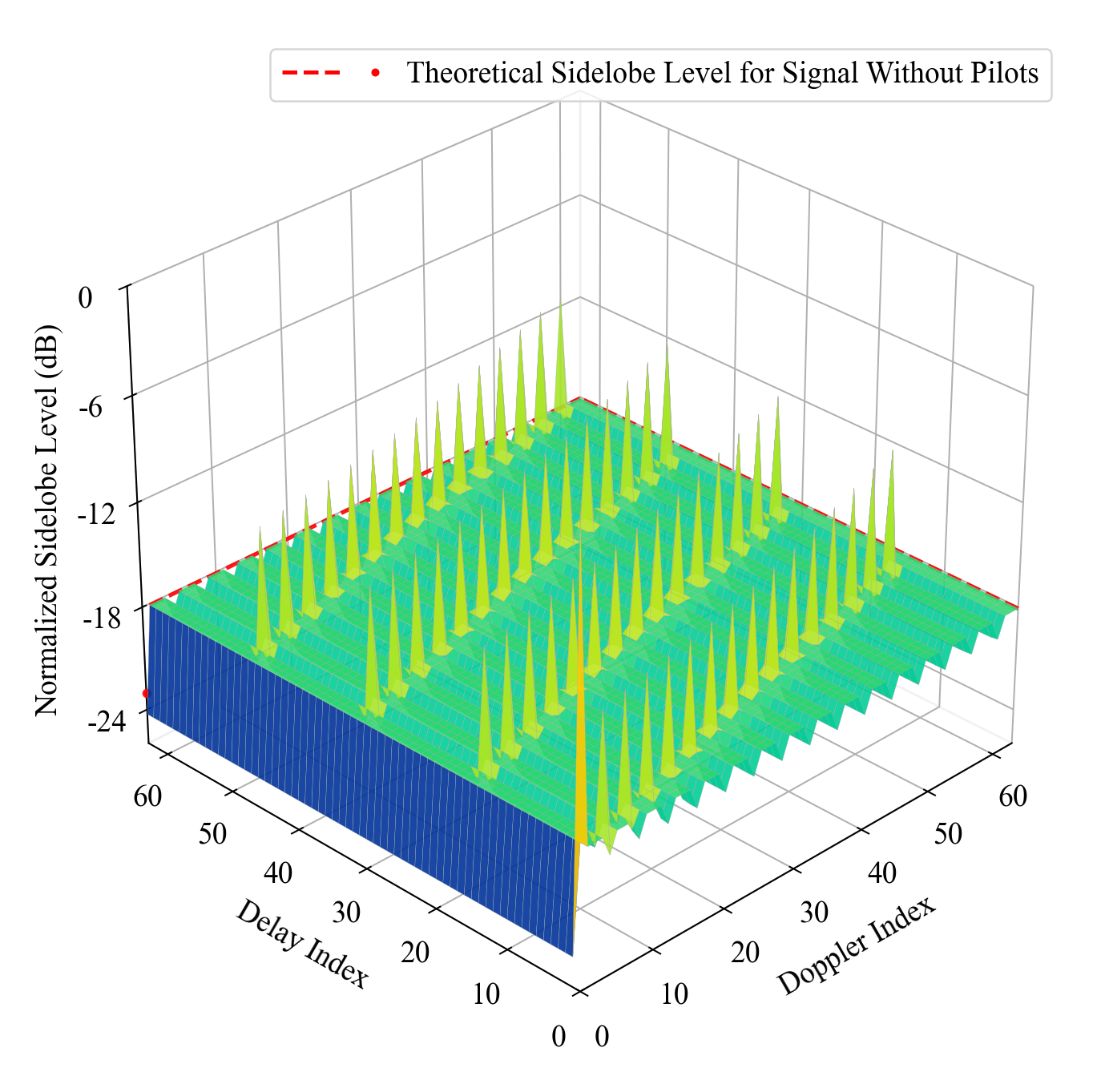}
		\caption{Comb-type Pilot, Theoretical}
		\label{chutian3}%文中引用该图片代号
	\end{subfigure}
	\centering
	\begin{subfigure}{0.48\linewidth}
		\centering
		\includegraphics[width=1\linewidth]{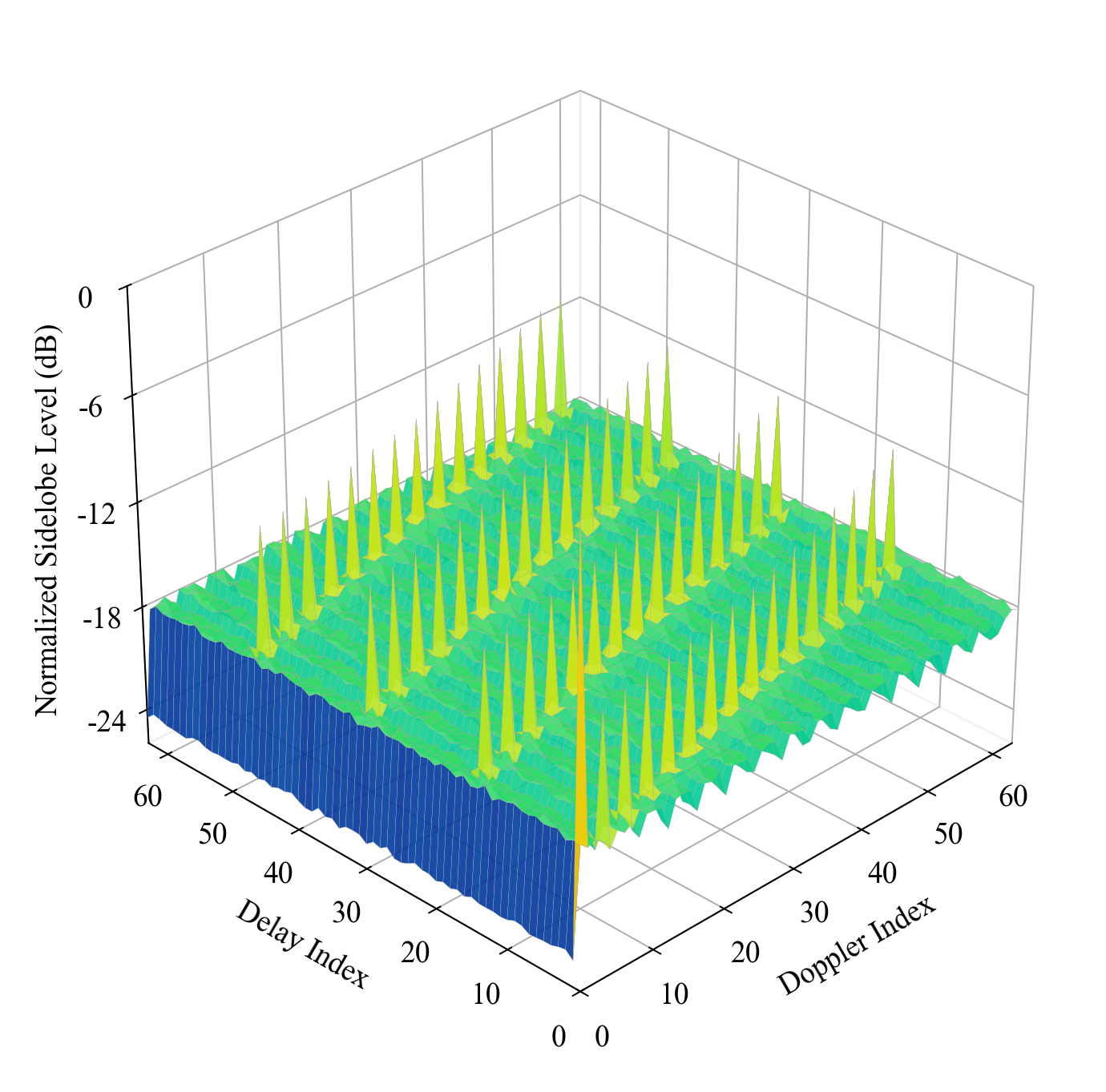}
		\caption{Comb-type Pilot, Simulated}
		\label{chutian3}%文中引用该图片代号
	\end{subfigure}
	\caption{The mean squared DP-AF of OFDM signals with/without pilots, where $N=64$, $L=16$. ZC sequences are adopted as pilots, with block-type and comb-type pattern.} 
	\label{fig3}
\end{figure}

\subsection{Effect of Pilots on the DP-AF of 1D OFDM signals} % ZC2, ZC3
For the configuration of pilot signals, we adopt Zadoff–Chu (ZC) sequence \cite{ZC}, which is widely used as pilot symbols in OFDM communication systems due to their favorable autocorrelation properties. The ZC sequence is expressed as $z_l=e^{-j\frac{\pi u l(l+1)}{L_{ZC}}},\ 0\leq l\leq L-1$. In the simulations of this work, the root of ZC sequence $u$ is set to $1$. After the $L$ subcarrier indices allocated to pilot signals are determined as $0\leq i_{0}<i_{1}\ldots<i_{L-1}\leq N-1$, the mapping configuration is defined as $z_l=s_{i_l}=p_{i_l},\ \forall l$. As shown in Fig. 1, We examine two pilot patterns: one places pilots at equal intervals on subcarriers, and the other concentrates pilots over a block of consecutive subcarriers. In this work, the former is referred to as the comb‑type pattern, while the latter is termed the block‑type pattern. 

As observed in Fig. 2a, for a 1D OFDM signal composed entirely of random communication symbols, the sidelobe level of its DP‑AF is binary valued. The zero-Doppler cut exhibits low sidelobes, which is consistent with the findings reported in \cite{zhang2025discrete}. After embedding the ZC sequences, it can be seen in Fig. 2c and Fig. 2e that the inserted unit-modulus pilots can reduce the sidelobe level in certain regions compared with the case without pilots, at the cost of raising the sidelobe level in other regions. Specifically, the comb-type pattern can introduce periodic peaks and troughs along the delay and Doppler dimensions. In contrast, the block-type pattern can produce densely distributed high sidelobes in some areas and densely distributed low sidelobes in others. 

When ZC sequences are used as pilot symbols, these different pilot patterns yield distinct characteristics of DP-AF. Furthermore, these observations suggest that, through careful design of the pilot pattern, it may be possible to modify the characteristics of the ambiguity function, thereby meeting specific sensing requirements.

\subsection{Effect of Pilots on the FST-AF of 2D OFDM signals}
As observed in Fig. 3, under the FST-AF formulation, the levels of sidelobe equal to the same value. After embedding $L$ unit-modulus pilots at arbitrary positions and with arbitrary symbols, the sidelobe level will decrease by a certain amount. Combined with the results in \eqref{eq:expect_fst}, this indicates that for a 2D random OFDM signal under the FST representation, its mean squared FST-AF under the small-Doppler approximation depends exclusively on the number of the unit-modulus pilots. The theoretical nature of FST-AF implies that, to further reveal the impact of the pilot symbols and pilot patterns on the actual sensing performance of 2D OFDM signals, it is necessary to go beyond the assumptions of the existing framework and adopt additional evaluation approaches.

\begin{figure}[htbp]
	\centering
	\begin{subfigure}{0.48\linewidth}
		\centering
		\includegraphics[width=1\linewidth]{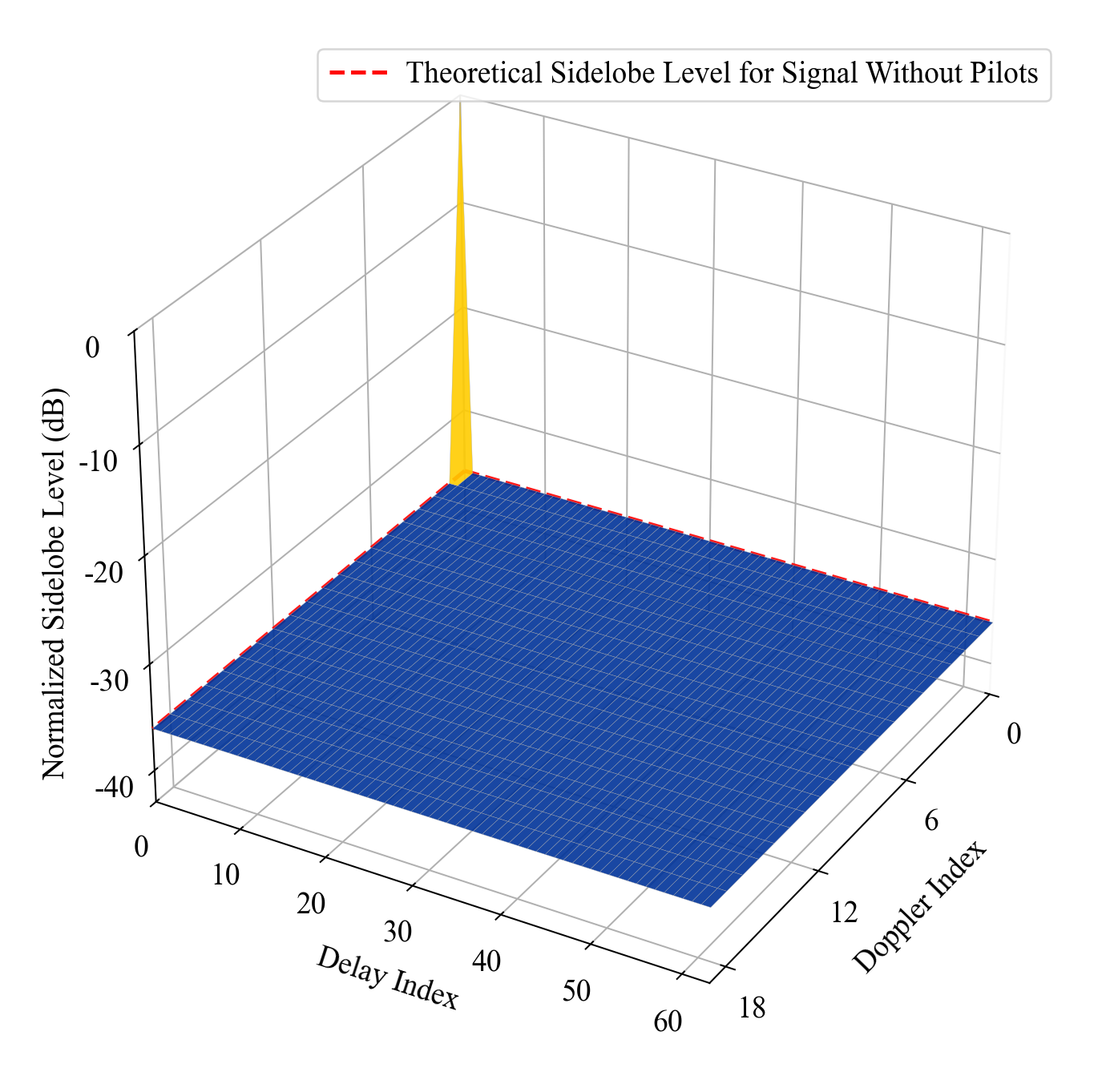}
		\caption{Without Pilot, Theoretical}
		\label{chutian3}%文中引用该图片代号
	\end{subfigure}
	\centering
	\begin{subfigure}{0.48\linewidth}
		\centering
		\includegraphics[width=1\linewidth]{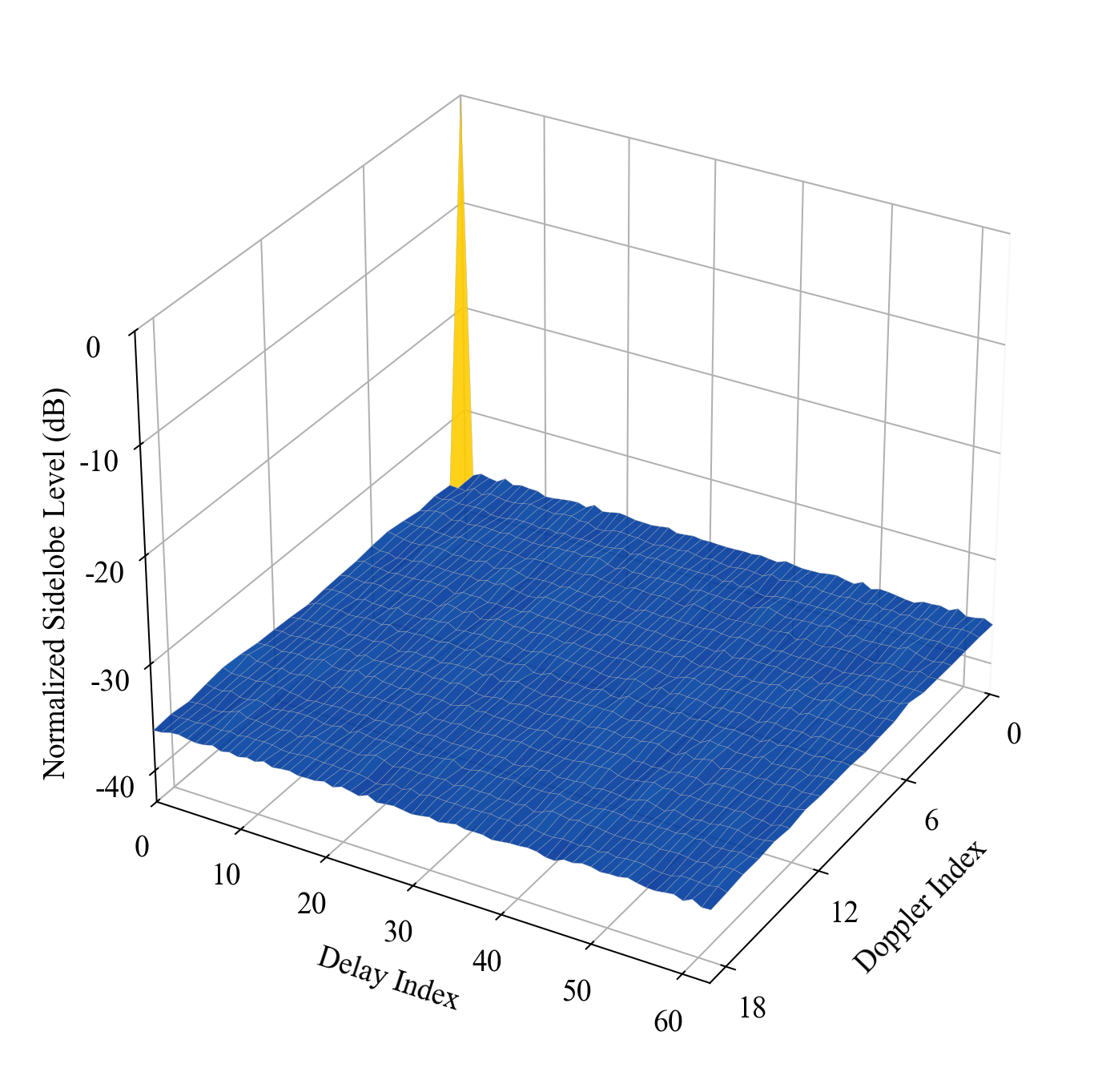}
		\caption{Without Pilot, Simulated}
		\label{chutian3}%文中引用该图片代号
	\end{subfigure}
	\centering
	\begin{subfigure}{0.48\linewidth}
		\centering
		\includegraphics[width=1\linewidth]{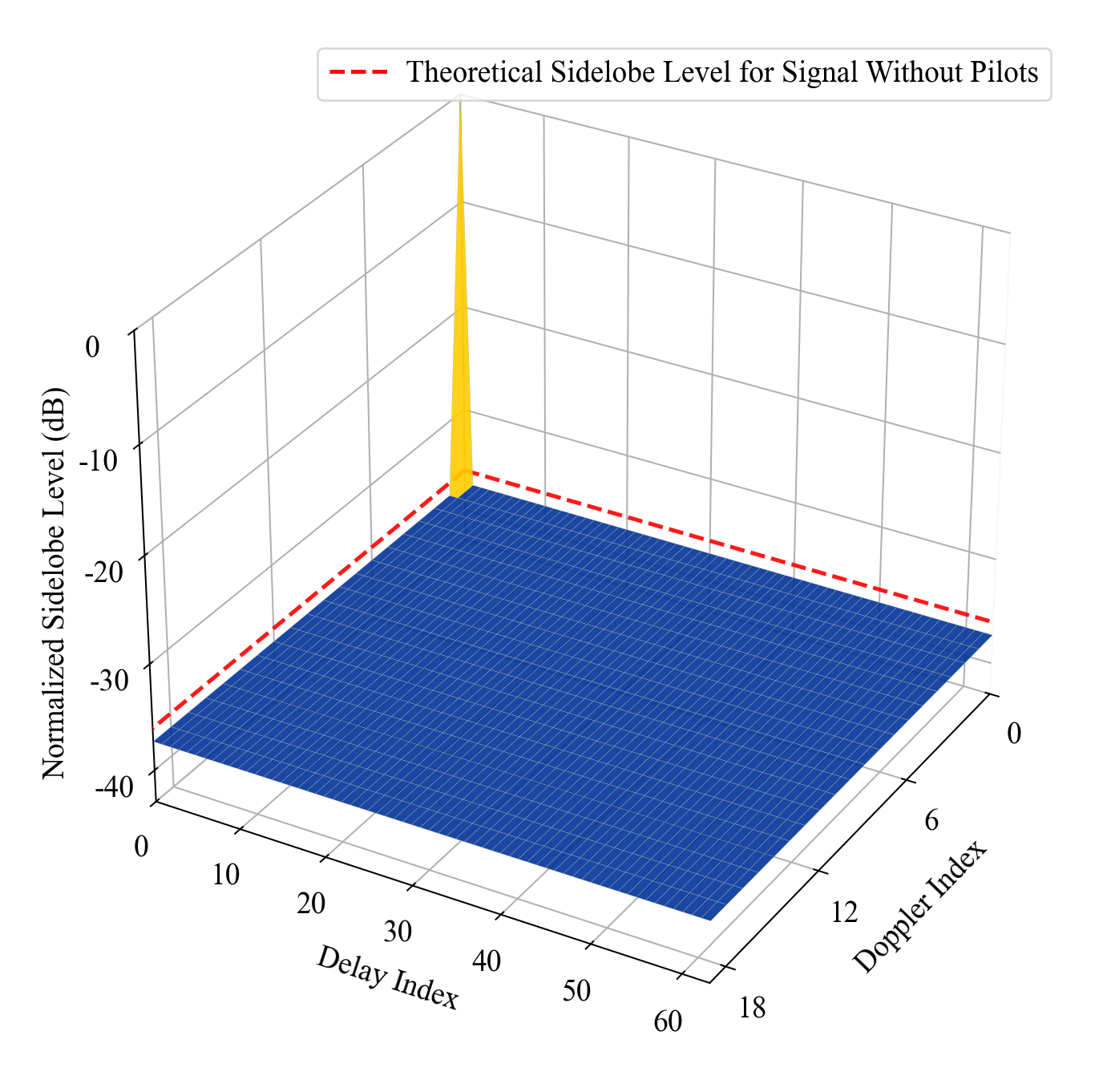}
		\caption{With Pilot, Theoretical}
		\label{chutian3}%文中引用该图片代号
	\end{subfigure}
	\centering
	\begin{subfigure}{0.48\linewidth}
		\centering
		\includegraphics[width=1\linewidth]{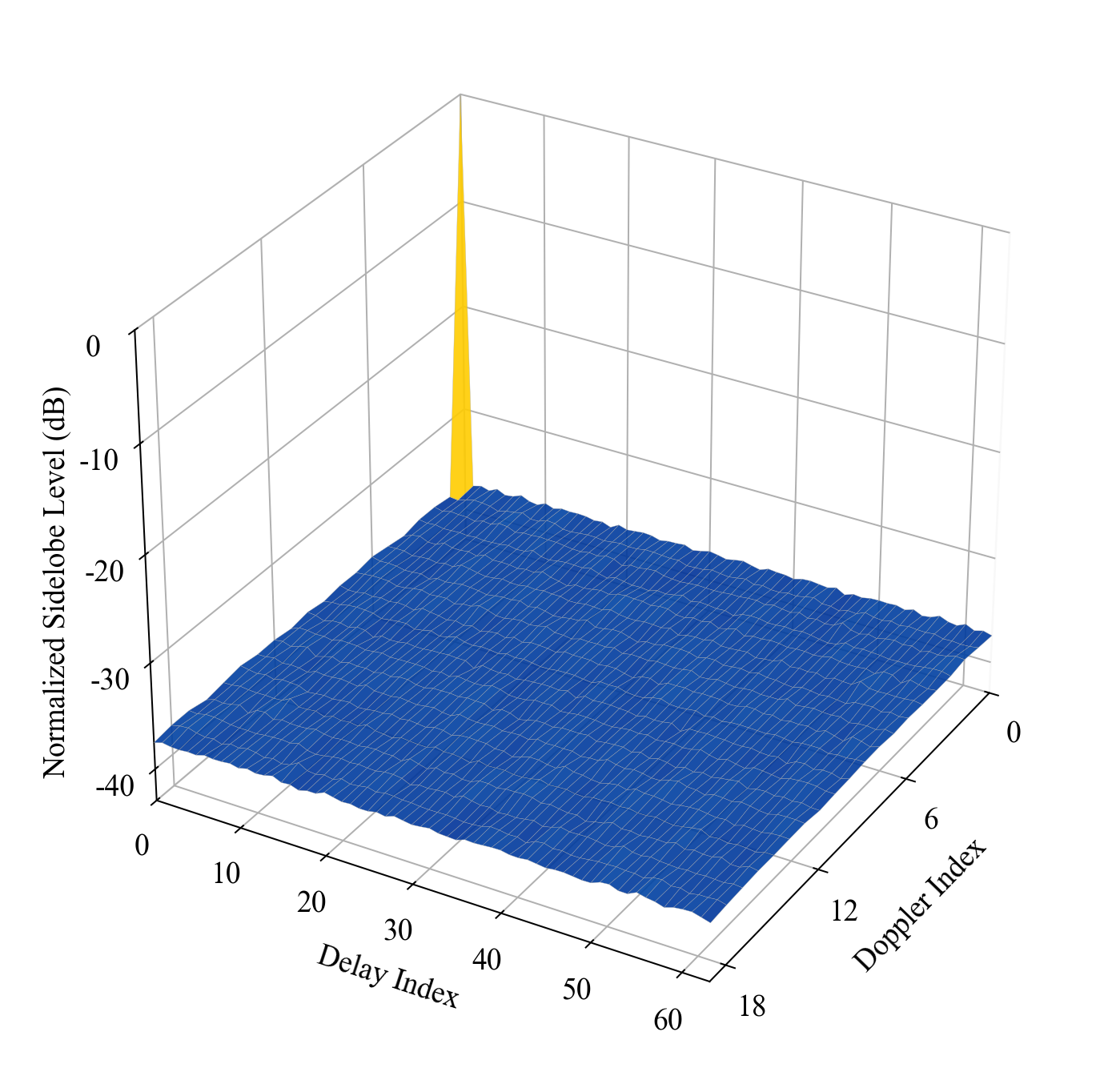}
		\caption{With Pilot, Simulated}
		\label{chutian3}%文中引用该图片代号
	\end{subfigure}
	\caption{The mean squared FST-AF of OFDM signals with/without pilots, where $N=64$, $M=20$, $L=16M$.} 
	\label{fig3}
\end{figure}

\section{Conclusions}
This study investigated the statistical ambiguity characteristics of random OFDM signals with embedded unit-modulus pilots. Analytical formulas for the mean squared DP-AF and FST-AF were established. Results demonstrate that under a fixed signal length and constellation scheme, the mean squared DP-AF is shaped by pilot patterns, pilot symbols and the number of pilots, while the mean squared FST-AF is determined exclusively by the number of pilots. Simulations confirm the theoretical derivations. 

\section*{Appendix A \\ Proof of Proposition 1}
Let $\mathbf{A}=\mathbf{F}_N\mathbf{D}_{N,q}\mathbf{J}_{N,k}\mathbf{F}_N^H$. Its entries are given by
\begin{equation}\begin{aligned}\label{eq:aentry}a_{l,r}=e^{-j2\pi rk/N}\delta_{r,\langle l-q\rangle_N}.\end{aligned}\end{equation}

According to \eqref{eq:def_dp}, the mean squared DP-AF is given by 
\begin{equation}\begin{aligned}\label{eq:expand_expect_dp}\mathbb{E}\left(|\mathcal{A}_{\mathrm{DP}}(k,q)|^2\right)&=\mathbb{E}\left(\left|\mathbf{x}^H\mathbf{D}_{N,q}\mathbf{J}_{N,k}\mathbf{x}\right|^2\right)=\mathbb{E}\left(\left|\mathbf{s}^H\mathbf{A}\mathbf{s}\right|^2\right)\\&=\sum_{l,r,m,n}a_{l,r}a_{m,n}^*\mathbb{E}\left(s_l^*s_rs_ms_n^*\right).
\end{aligned}\end{equation}
Substituting \eqref{eq:swpd} into $s_l^*s_rs_ms_n^*$ yields:
\begin{align}\label{eq:ssss}\nonumber &\ s_l^*s_rs_ms_n^*\\&=\nonumber\left[w_lp_l^*+(1-w_l)d_l^*\right] \cdot \left[w_rp_r+(1-w_r)d_r\right]\\&\ \cdot \left[w_mp_m+(1-w_m)d_m\right] \cdot \left[w_np_n^*+(1-w_n)d_n^*\right].
\end{align}
According to \eqref{eq:defdi}, $d_l$ is a random communication symbol if $w_l=0$, and conversely, $d_l=0$ if $w_l=1$. From the assumption in \eqref{eq:asp1} and \eqref{eq:asp2} we have
\begin{align}\label{eq:expect_d}
\mathbb{E}[(1-w_l)d_l^*]=\mathbb{E}[(1-w_r)d_r]=0,
\end{align}
\begin{align}\label{eq:expect_dd_1}
\mathbb{E}[(1-w_l)(1-w_r)d_l^*d_r]=(1-w_l)(1-w_r)\delta_{l,r},
\end{align}
\begin{align}\label{eq:expect_dd_2}
\mathbb{E}[(1-w_m)(1-w_r)d_md_r]=\mathbb{E}[(1-w_l)(1-w_n)d_l^*d_n^*]=0,
\end{align}
\begin{align}\label{eq:expect_ddd}
&\ \nonumber\mathbb{E}[(1-w_l)(1-w_r)(1-w_m)d_l^*d_rd_m]\\&=\mathbb{E}[(1-w_l)(1-w_r)(1-w_n)d_l^*d_rd_n^*]=0.
\end{align}
\begin{align}\label{eq:expect_dddd}
&\ \nonumber\mathbb{E}[(1-w_l)(1-w_r)(1-w_m)(1-w_n)d_l^*d_rd_md_n^*]\\&\nonumber=(1-w_l)(1-w_r)(1-w_m)(1-w_n)(\delta_{l,r}\delta_{m,n}+\delta_{l,m}\delta_{r,n})\\&+(1-w_l)(1-w_r)(1-w_m)(1-w_n)(\kappa-2)\delta_{l,r,m,n}.
\end{align}
By substituting \eqref{eq:expect_d}, \eqref{eq:expect_dd_1}, \eqref{eq:expect_dd_2}, \eqref{eq:expect_ddd} and \eqref{eq:expect_dddd} into the expectation of \eqref{eq:ssss}, we have:
\begin{equation}\begin{aligned}\label{eq:expect_ssss}&\ \mathbb{E}\left(s_l^*s_rs_ms_n^*\right)\\& = (w_lp_l^*w_rp_r)(w_mp_mw_np_n^*)\\&+(1-w_l)(1-w_r)(1-w_m)(1-w_n)(\delta_{l,r}\delta_{m,n}+\delta_{l,m}\delta_{r,n})\\&+(1-w_l)(1-w_r)(1-w_m)(1-w_n)(\kappa-2)\delta_{l,r,m,n}
\\&+(1-w_l)(1-w_r)(w_mp_mw_np_n^*)\delta_{l,r}
\\&+(1-w_l)(1-w_m)(w_rp_rw_np_n^*)\delta_{l,m}
\\&+(1-w_r)(1-w_n)(w_lp_l^*w_mp_m)\delta_{r,n}
\\&+(1-w_m)(1-w_n)(w_lp_l^*w_rp_r)\delta_{m,n}.
\end{aligned}\end{equation}
By substituting \eqref{eq:expect_ssss} into \eqref{eq:expand_expect_dp}, it follows that
\begin{align}\label{eq:expect_dp_witha}
&\ \nonumber\mathbb{E}\left(|\mathcal{A}_{\mathrm{DP}}(k,q)|^2\right)\\&\nonumber=\sum_{l,r}a_{l,r}(w_lp_l^*w_rp_r)\sum_{m,n}a_{m,n}^*(w_mp_mw_np_n^*)\\&\nonumber+\sum_{l,m}a_{l,l}a_{m,m}^*(1-w_l)^2(1-w_m)^2\\&\nonumber+\sum_{l,r}a_{l,r}a_{l,r}^*(1-w_l)^2(1-w_r)^2\\&\nonumber+\sum_{l}a_{l,l}a_{l,l}^*(1-w_l)^4(\kappa-2)\\&\nonumber+\sum_{l}a_{l,l}(1-w_l)^2\sum_{m,n}a_{m,n}^*(w_mp_mw_np_n^*)\\&\nonumber+\sum_{l}(1-w_l)^2\sum_{r}a_{l,r}w_rp_r\sum_{n}a_{l,n}^*w_np_n^*\\&\nonumber+\sum_{r}(1-w_r)^2\sum_{l}a_{l,r}w_lp_l^*\sum_{m}a_{m,r}^*w_mp_m\\&+\sum_{m}a_{m,m}^*(1-w_m)^2\sum_{l,r}a_{l,r}(w_lp_l^*w_rp_r).
\end{align}
By substituting \eqref{eq:aentry} into \eqref{eq:expect_dp_witha}, we arrive at
\begin{equation}\begin{aligned}
&\ \mathbb{E}\left(|\mathcal{A}_{\mathrm{DP}}(k,q)|^2\right)\\&=\left|\sum_{l}p_lp_{\langle l-q\rangle_N}^*\cdot e^{j2\pi lk/N}\right|^2\\&+\sum_l(1-w_l)^2(1-w_{\langle l-q\rangle_N})^2\\&+\sum_l(1-w_l)^2\left(w_{\langle l-q\rangle_N}^2+w^2_{\langle l+q\rangle_N}\right)\\&+\left|\sum_{l}(1-w_l)^2\cdot e^{j2\pi lk/N}\right|^2\delta_{q,0}+(\kappa-2)\sum_l(1-w_l)^4\delta_{q,0}\\&+\left[\sum_{l}(1-w_l)^2\cdot e^{-j2\pi lk/N}\right]\left[\sum_{m}(w_m)^2\cdot e^{j2\pi mk/N}\right]
\delta_{q,0}\\&+\left[\sum_{m}(1-w_m)^2\cdot e^{j2\pi mk/N}\right]\left[\sum_{l}(w_l)^2\cdot e^{-j2\pi lk/N}\right]
\delta_{q,0},
\end{aligned}\end{equation}
\iffalse\begin{equation}\begin{aligned}
&\ \mathbb{E}\left(|\mathcal{A}_{\mathrm{DP}}(k,q)|^2\right)\\&=\left|\sum_{l}p_lp_{\langle l-q\rangle_N}^*\cdot e^{j2\pi lk/N}\right|^2\\&+\sum_l(1-w_l)(1+w_{\langle l+q\rangle_N})\\&+\left|\sum_{l}(1-w_l)\cdot e^{j2\pi lk/N}\right|^2\delta_{q,0}+(\kappa-2)\sum_l(1-w_l)\delta_{q,0}\\&+\left[\sum_{l}(1-w_l)\cdot e^{-j2\pi lk/N}\right]\left[\sum_{m}w_m\cdot e^{j2\pi mk/N}\right]
\delta_{q,0}\\&+\left[\sum_{m}(1-w_m)\cdot e^{j2\pi mk/N}\right]\left[\sum_{l}w_l\cdot e^{-j2\pi lk/N}\right]
\delta_{q,0},
\end{aligned}\end{equation}\fi
which is equivalent to the expression in Proposition 1.
%\hfill $\square$

\section*{Appendix B \\ Proof of Proposition 2}
Let $\mathbf{F}_N^H=\mathbf{G}$,  $\mathbf{F}_M=\mathbf{H}$, from \eqref{eq:fst} we have:
\begin{align}
\mathcal{A}_{\mathrm{FST}}(k,q)&=\sqrt{MN}\sum_{n=0}^{N-1}\sum_{m=0}^{M-1}g_{k,n}h_{m,q}\left|s_{n,m}\right|^2,
\end{align}
then we have
\iffalse
\begin{align}
&\nonumber|\mathcal{A}_{\mathrm{FST}}(k,q)|^2 = \sum_{m=1}^{N}\sum_{n=1}^{M} \left|s_{m,n}\right|^4\\&+MN\left(\sum_{(m,n)\neq(m',n')}g_{k,m}g_{k,m'}^*h_{n,q}h_{n',q}^*\left|s_{m,n}\right|^2\left|s_{m',n'}\right|^2\right).
\end{align}
\fi
\begin{align}
&\nonumber\ |\mathcal{A}_{\mathrm{FST}}(k,q)|^2 = \sum_{n=0}^{N-1}\sum_{m=0}^{M-1} \left|s_{n,m}\right|^4\\&+MN\sum_{(n,m)\neq(l,r)}g_{k,n}g_{k,l}^*h_{m,q}h_{r,q}^*\left|s_{n,m}\right|^2\left|s_{l,r}\right|^2.
\end{align}
It follows that
\iffalse
\begin{align}
&\nonumber\ \mathbb{E}\left(|\mathcal{A}_{\mathrm{FST}}(k,q)|^2\right)\nonumber= L+(MN-L)\cdot\kappa\\&\nonumber+MN\sum_{(m,n)\neq(m',n')}g_{k,m}g_{k,m'}^*h_{n,q}h_{n',q}^*
\\&\nonumber=L+(MN-L)\cdot\kappa-MN\\&+\sum_{m,n,m',n'}\text{cos}\left(\frac{2\pi}{N}(m-m')(k-1)-\frac{2\pi}{M}(n-n')(q-1)\right),
\end{align}
\fi
\begin{align}
&\nonumber\ \mathbb{E}\left(|\mathcal{A}_{\mathrm{FST}}(k,q)|^2\right)\nonumber\\&= L+(MN-L)\cdot\kappa\nonumber+MN\sum_{(n,m)\neq(l,r)}g_{k,n}g_{k,l}^*h_{m,q}h_{r,q}^*
\\&\nonumber=L+(MN-L)\cdot\kappa-MN\\&+\sum_{n,m,l,r}\text{cos}\left[\frac{2\pi k}{N}(n-l)-\frac{2\pi q}{M}(m-r)\right],
\end{align}
which is equivalent to the expression in Proposition 2.
%\hfill $\square$

\bibliographystyle{IEEEtran}
\bibliography{reference}

@ARTICLE{Liu2026JSAC,
  author={Liu, Fan and Liu, Ya-Feng and Cui, Yuanhao and Masouros, Christos and Xu, Jie and Xiao Han, Tony and Buzzi, Stefano and Eldar, Yonina C. and Jin, Shi},
  journal={IEEE J. Sel. Areas Commun.}, 
  title={Sensing With Communication Signals: From Information Theory to Signal Processing}, 
  year={2026},
  volume={44},
  number={},
  pages={1-30},
  doi={10.1109/JSAC.2025.3614025}}

@ARTICLE{Both_Pilots_and_Data_Payloads,
  author={Xu, Chen and Yu, Xianghao and Liu, Fan and Jin, Shi},
  journal={IEEE Trans. Wireless Commun.}, 
  title={Exploiting Both Pilots and Data Payloads for Integrated Sensing and Communications}, 
  year={2026},
  volume={25},
  number={},
  pages={5573-5586},
  doi={10.1109/TWC.2025.3619280}}

@ARTICLE{CP-OFDM,
  author={Liu, Fan and Zhang, Ying and Xiong, Yifeng and Li, Shuangyang and Yuan, Weijie and Gao, Feifei and Jin, Shi and Caire, Giuseppe},
  journal={IEEE Trans. Inf. Theory}, 
  title={{CP-OFDM} Achieves the Lowest Average Ranging Sidelobe Under {QAM/PSK} Constellations}, 
  year={2025},
  volume={71},
  number={9},
  pages={6950-6967},
  doi={10.1109/TIT.2025.3591267}}

@ARTICLE{iceberg,
  author={Liu, Fan and Xiong, Yifeng and Lu, Shihang and Li, Shuangyang and Yuan, Weijie and Masouros, Christos and Jin, Shi and Caire, Giuseppe},
  journal={IEEE Trans. Signal Process.}, 
  title={Uncovering the Iceberg in the Sea: Fundamentals of Pulse Shaping and Modulation Design for Random {ISAC} Signals}, 
  year={2025},
  volume={73},
  number={},
  pages={2511-2526},
  doi={10.1109/TSP.2025.3580596}}

@ARTICLE{6GBeyond,
  author={Liu, Fan and Cui, Yuanhao and Masouros, Christos and Xu, Jie and Han, Tony Xiao and Eldar, Yonina C. and Buzzi, Stefano},
  journal={IEEE J. Sel. Areas Commun.}, 
  title={Integrated Sensing and Communications: Toward Dual-Functional Wireless Networks for {6G} and Beyond}, 
  year={2022},
  volume={40},
  number={6},
  pages={1728-1767},
  doi={10.1109/JSAC.2022.3156632}}

@article{zhang2025discrete,
  title={On Discrete Ambiguity Functions of Random Communication Waveforms},
  author={Zhang, Ying and Liu, Fan and Xiong, Yifeng and Yuan, Weijie and Li, Shuangyang and Zheng, Le and Han, Tony Xiao and Masouros, Christos and Jin, Shi},
  journal={arXiv preprint arXiv:2512.08352},
  year={2025}
}

@ARTICLE{ZC,
  author={Wei, Zhiqing and Mei, Dongyang and Wang, Lin and Zan, Jiaqi and Dong, Zixian and Feng, Zhiyong},
  journal={IEEE Trans. Veh. Technol.}, 
  title={Coprime Sensing Reference Signal Design Based on {Zadoff–Chu} Sequence and Index Modulation for Uplink {ISAC} System}, 
  year={2026},
  volume={75},
  number={2},
  pages={2777-2790},
  doi={10.1109/TVT.2025.3604403}}

@ARTICLE{Over_the_Years,
  author={Zhang, Di and Cui, Yuanhao and Cao, Xiaowen and Su, Nanchi and Gong, Yi and Liu, Fan and Yuan, Weijie and Jing, Xiaojun and Andrew Zhang, J. and Xu, Jie and Masouros, Christos and Niyato, Dusit and Di Renzo, Marco},
  journal={IEEE Commun. Surveys Tuts.}, 
  title={Integrated Sensing and Communications Over the Years: An Evolution Perspective}, 
  year={2026},
  volume={28},
  number={},
  pages={5014-5048},
  doi={10.1109/COMST.2026.3655674}}

@INPROCEEDINGS{RSWCNC,
  author={Khosroshahi, Keivan and Sehier, Philippe and Mekki, Sami and Suppa, Michael},
  booktitle={IEEE Wireless Commun. Netw. Conf. (WCNC)}, 
  title={Localization Accuracy Improvement in Multistatic {ISAC} with {LoS/NLoS} Condition Using {5G NR} Signals}, 
  year={2025},
  volume={},
  number={},
  pages={1-6},
  doi={10.1109/WCNC61545.2025.10978602}}

@ARTICLE{AndrewAF,
  author={Ni, Zhitong and Andrew Zhang, J. and Ping Liu, Ren and Lee, Inkyu},
  journal={IEEE Wireless Commun. Lett.}, 
  title={Ambiguity Function-Based Filter Designs for {OFDM ISAC} Systems}, 
  year={2026},
  volume={15},
  number={},
  pages={181-185},
  doi={10.1109/LWC.2025.3623243}}

\end{document}